\documentclass[%
 aip,
 amsmath,amssymb,
 reprint,%
]{revtex4-1}

\usepackage{graphicx}
\usepackage{dcolumn}
\usepackage{bm}

\usepackage[utf8]{inputenc}
\usepackage[T1]{fontenc}
\usepackage{mathptmx}
\usepackage{etoolbox}
\usepackage{hyperref}

\makeatletter
\def\@email#1#2{%
 \endgroup
 \patchcmd{\titleblock@produce}
  {\frontmatter@RRAPformat}
  {\frontmatter@RRAPformat{\produce@RRAP{*#1\href{mailto:#2}{#2}}}\frontmatter@RRAPformat}
  {}{}
}%
\makeatother
\begin{document}

\preprint{AIP/123-QED}

\title{Dominant scattering mechanisms in the low/high electric field transport in cryogenic 2D confinement in silicon (110) with high-$\kappa$ oxides}
\author{Hsin-Wen Huang}
\author{Xi-Jun Fang}
\affiliation{ 
Graduate Institute of Photonics and Optoelectronics and Department of Electrical Engineering, National Taiwan University, Sec.4, Roosevelt Rd., Taipei 10617, Taiwan.
}

\author{Edward Chen}
\affiliation{ 
Corporate Research, Taiwan Semiconductor Manufacturing Company Ltd., 168, Kehuan Rd., Baoshan Township, Hsinchu County 308-001, Taiwan.
}%

\author{Yuh-Renn Wu}
\email{yrwu@ntu.edu.tw}
\affiliation{ 
Graduate Institute of Photonics and Optoelectronics and Department of Electrical Engineering, National Taiwan University, Sec.4, Roosevelt Rd., Taipei 10617, Taiwan.
}
\affiliation{ 
Center for Quantum Science and Engineering, 1, Roosevelt Rd. Sec.4,, Taipei 10617, Taiwan.
}

\date{July 2026}

\begin{abstract}
The performance of silicon nano-devices at cryogenic temperatures is critical for quantum qubit control circuits and space applications. Using multi-valley Monte Carlo simulations, we investigate electron transport in Si~(110) systems. At low electric fields, phonon absorption becomes negligible, and mobility is governed by competition between remote Coulomb scattering~(RCS) at low inversion charge density and surface roughness scattering~(SRS) at high density, leading to a mobility peak. High-$\kappa$ dielectrics such as $\mathrm{HfO_2}$ introduce remote phonon scattering~(RPS), which suppresses mobility. Under high electric fields, phonon emission dominates at 4K, limiting velocity enhancement and resulting in limited current improvement.
\end{abstract}

\maketitle

\section{Introduction}
Cryogenic metal oxide semiconductor field-effect transistors (MOSFETs) are essential for peripheries of a quantum computer and are also relevant to deep-space and satellite electronic applications \cite{RN1,RN20,RN19,RN29}. Understanding two-dimensional (2D) electron transport under silicon (110) confinement is crucial for advancing cryogenic FinFET and other nonplanar silicon technologies, where carrier transport strongly depends on surface orientation and body thickness \cite{RN24,RN30}. Previous studies have investigated the temperature dependence of electron mobility in bulk MOSFETs under (100) and (110) field-induced quantum confinement with $\mathrm{SiO_2}$ and $\mathrm{HfO_2}$ gate dielectrics \cite{RN2,RN3,RN4}. Furthermore, cryogenic operation of FinFET devices has been explored in several experimental and modeling studies \cite{RN25,RN26,RN27}. However, detailed investigations of electron transport under Si(110) confinement in advanced FinFET geometries remain limited. Moreover, the field-dependent velocity is important due to the short channel lengths in the tens of nanometers \cite{RN23}. This work employs multi-subband Monte Carlo simulation to investigate electron mobility in silicon (110) confinement under cryogenic conditions. By incorporating quantum confinement effects and solving the Poisson-Schr$\mathrm{\ddot{o}}$dinger equations, the dominant scattering rates and their impact on transport properties have been analyzed. Our results reveal that surface roughness and Coulomb scattering dominate at low temperatures, while high-$\kappa$ dielectrics introduce additional mobility-limiting factors. These findings provide valuable insights for cryogenic semiconductor design and nanoelectronic applications.

\section{Methodology}
\begin{figure}
\includegraphics[width=\linewidth, keepaspectratio]{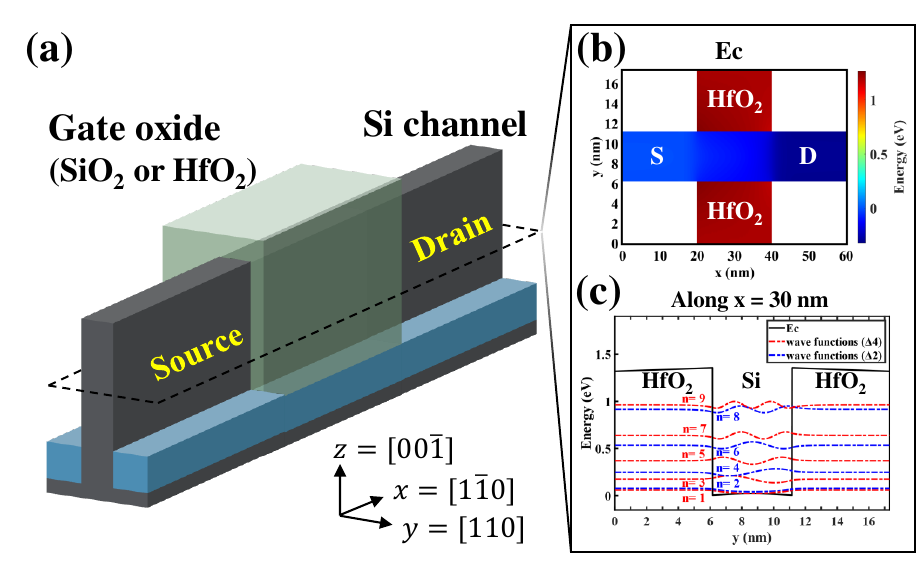}
\caption{\label{Fig:fig1_ppt_to_pdf_cropped} (a) The FinFET structure used in this study. The quantum confinement is in the (110) direction. (b) shows the Ec of the channel at the cross section. (c) shows the conduction band profile and the calculated subband energies, together with their corresponding wave functions, obtained from the Poisson and Schr$\mathrm{\ddot{o}}$dinger solver along the cross section shown in (b). }
\end{figure}

This study focuses on electron mobility in silicon (110) confinement systems under cryogenic conditions. Figure \ref{Fig:fig1_ppt_to_pdf_cropped} shows the FinFET structure \cite{RN21} studied in this paper. As shown in Fig. \ref{Fig:fig1_ppt_to_pdf_cropped}, the lateral confinement at each position is calculated, and the corresponding wave functions are used for the scattering rate evaluation. To model carrier transport, we employed a multi-valley Monte Carlo method. 
To account for quantum confinement under Si(110), the X-valley states were explicitly separated into the $\Delta_4$ and $\Delta_2$ subbands, and the corresponding quantization effective masses were used for the Schr$\mathrm{\ddot{o}}$dinger subband calculation~\cite{RN5}.

Then, the Poisson and Schr$\mathrm{\ddot{o}}$dinger equations were solved to acquire the wave function as shown in Fig.\ref{Fig:fig1_ppt_to_pdf_cropped}. In particular, we apply the localized landscape model \cite{RN6} to obtain the effective quantum potential, which is easier to couple with the Poisson and drift-diffusion solver, and we can obtain a better-converged result. Further, these wave functions were then utilized to calculate scattering rates for various mechanisms, including acoustic phonon scattering (APS), optical phonon scattering (OPS), remote phonon scattering (RPS), surface roughness scattering (SRS), remote Coulomb scattering (RCS), and ionized impurity scattering (IIS). 
Representative theoretical treatments and parameter sources for these scattering mechanisms can be found in Refs.~\cite{RN7,RN8,RN9,RN10,RN11,RN12,RN31,RN13,RN14,RN22,RN33}. The specific theoretical model adopted for each scattering mechanism, together with the corresponding assumptions, parameter values, and literature sources, is summarized in the Supplementary Information and Table~S1.
To compute the scattering rates, we applied Fermi's Golden Rule \cite{RN15,RN16}, which is given by:

\begin{equation}
 W(k,k')=\frac{2\pi}{\hbar}\vert{M_{kk'}}\vert^2\delta(E_{k'}-E_{k})
 \end{equation}

In this equation, W(k,k$^{'}$) represents the scattering rate for a transition from the initial state k to the final state k$^{'}$, $M_{kk^{'}}$ denotes the matrix element of the scattering process, and $\delta(E_{k^{'}}-E_{k})$ means the energy conservation.

The Monte Carlo \cite{RN17,RN28} simulation tracks individual carriers through random sampling of scattering events, accounting for interactions with phonons, ionized impurities, and the surrounding material. This approach provides a detailed view of carrier dynamics, capturing the effects of complex interactions, particularly at low temperatures where phonon-related scattering diminishes and other mechanisms, like surface roughness and Coulomb scattering, become dominant.

The Monte Carlo scattering models are summarized in the Supplementary Information, and the complete parameter set used in the simulations is listed in Table~S1. In this work, the fixed-charge concentrations used in the remote Coulomb scattering model and the surface roughness parameters are treated as empirical fitting parameters, whereas the ionized impurity concentration is treated as an assumed device parameter. 

\section{Results and Discussion}
\begin{figure}
\includegraphics[width=\linewidth, keepaspectratio]{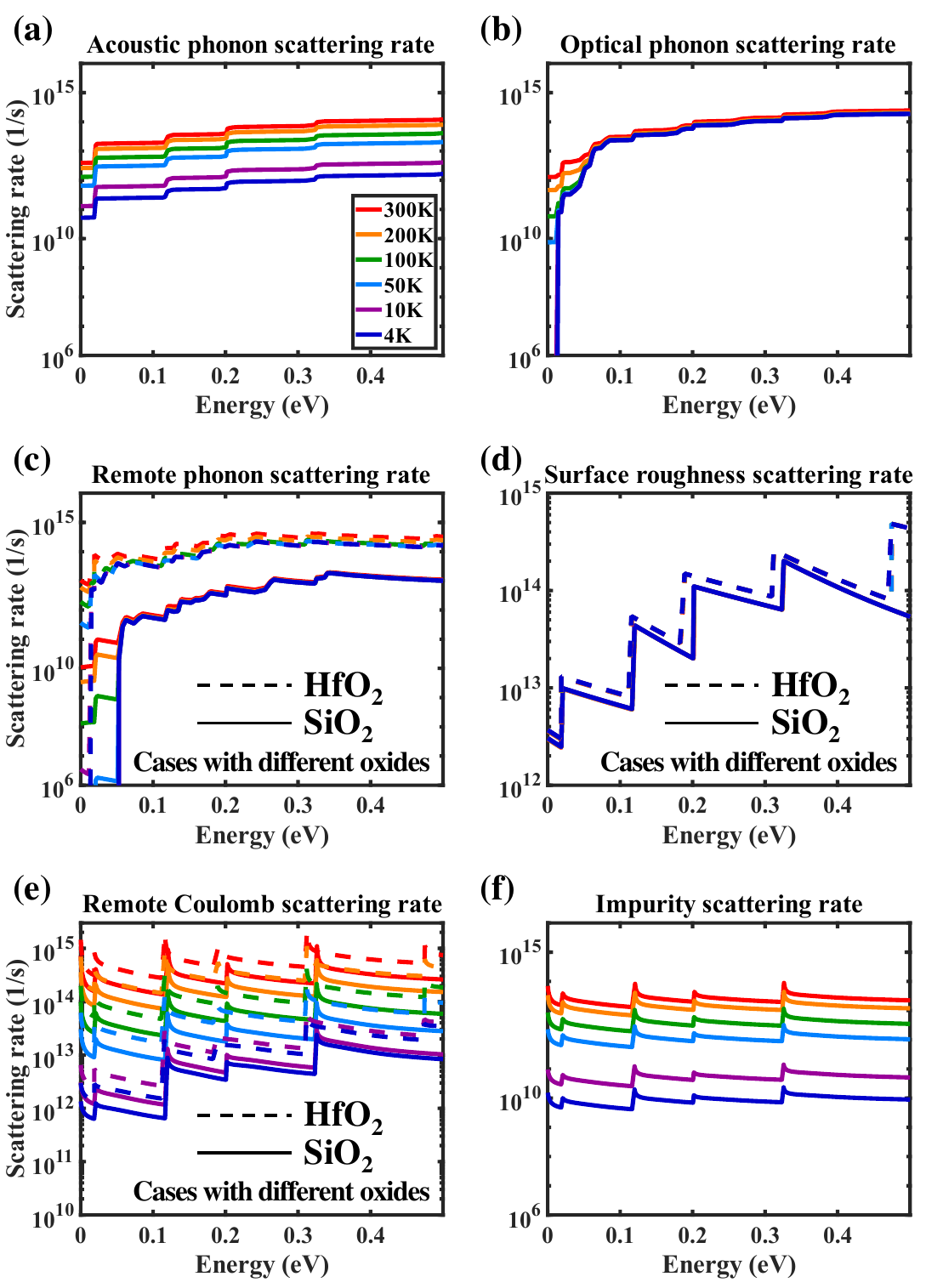}
\caption{\label{Fig:fig2_ppt_to_pdf_cropped} Scattering rates with electron energy. (a) Acoustic phonon. (b) Optical phonon. Intravalley and inter-valley transitions are included. 
(c) Remote phonon. 
(d) Surface roughness. In (c) and (d), the dashed and solid lines correspond to the $\mathrm{HfO_2}$ and $\mathrm{SiO_2}$ cases, respectively. 
(e) Remote Coulomb. 
(f) Impurity. 
The empirical SRS and RCS parameters and the assumed IIS concentration are described in the text and summarized in Table~S1.}
\end{figure}

Fig. \ref{Fig:fig2_ppt_to_pdf_cropped} shows the scattering rates of every scattering mechanism in our model with different temperatures. As shown in Fig. \ref{Fig:fig2_ppt_to_pdf_cropped}, phonon-caused scattering is crucial when discussing the electron transport property. Since the phonon is oriented from the thermal fluctuation, the electron-phonon scattering rate is highly related to the environment temperature. Fig. \ref{Fig:fig2_ppt_to_pdf_cropped}(a) is the acoustic phonon scattering rates of the silicon (110) confinement system. The detailed parameters used in the calculation are from Refs.~\cite{RN13,RN18}. The device in Ref.~\cite{RN18} is like a tri-gates structure, and the deformation potential the researcher used is 6.27 eV. However, in terms of SOI, Esseni found the value should be 14.6 eV \cite{RN13}, and this condition is more similar to our study. Therefore, 14.6 eV is chosen as the deformation potential in our calculation. As a result, the scattering rate decreases as the temperature decreases. Fig. \ref{Fig:fig2_ppt_to_pdf_cropped}(b) is the optical phonon scattering rates, including the intravalley and equivalent inter-valleys transitions (f and g scattering) by emitting or absorbing an optical phonon energy. As the temperature decreases, the electron optical phonon emission scattering dominates. As the temperature is below 50K, the electron-phonon emission scattering is directly related to carrier energy instead of phonon occupations. The phonon absorption scattering is diminished at low K.

For better gate control in the channel, high-$\kappa$ materials such as $\mathrm{HfO_2}$ are applied as the dielectric layer. However, these materials can introduce remote phonon scattering, which has been reported to degrade carrier mobility \cite{RN9}. In this study, we discuss the remote phonon scattering effect by comparing the conditions of two different oxide materials ($\mathrm{SiO_2}$ and $\mathrm{HfO_2}$) as the dielectric. Fig. \ref{Fig:fig2_ppt_to_pdf_cropped}(c) represents the remote phonon scattering rates. The parameters used in remote phonon scattering rate calculation are from Ref.\cite{RN9}. From the FinFET structure, we can simplify our simulation into the double gates condition. That is, the top and bottom dielectrics sandwich the silicon channel. The material of both sides of the oxide is the same. The width of the channel is set as 5 nm, and the thickness of the oxide is 1 nm. Compared with oxides in our $\mathrm{SiO_2}$ case, the $\mathrm{HfO_2}$ one has a much higher scattering rate. This is because of the large dielectric constant at low frequency.

Surface roughness scattering occurs when electrons move around the interface of two materials. The imperfection surface causes fluctuations in the potential energy landscape, leading to scattering events that affect the motion of the electrons, and this mechanism is independent of the temperature. 
In this work, the surface roughness parameters are treated as empirical fitting parameters, with $\lambda=1.0$~nm for both gate dielectrics, $\Delta=1.6$~nm for $\mathrm{SiO_2}$, and $\Delta=3.2$~nm for $\mathrm{HfO_2}$. Fig. \ref{Fig:fig2_ppt_to_pdf_cropped}(d) shows the surface roughness scattering rate.

Remote Coulomb scattering is induced from the charges located in the oxide or at the oxide-semiconductor interface. Due to the potential propagation of the remote Coulomb, the carrier's transport property in the channel is influenced. Since the Coulomb force is related to the distance between the oxide charges and channel carriers, knowing the distribution of the carriers in the channel is vital. 
Therefore, we chose a similar 2D electron gas density in the inversion layer ($\mathrm{n_{inv}}$) under temperatures (around $\mathrm{1.0 \times 10^{12}}$\,$\mathrm{cm^{-2}}$). In this work, the fixed-charge concentration is treated as an empirical fitting parameter and is set to $\mathrm{5.0 \times 10^{12}}$\,$\mathrm{cm^{-2}}$ for $\mathrm{SiO_2}$ and $\mathrm{1.0 \times 10^{13}}$\,$\mathrm{cm^{-2}}$ for $\mathrm{HfO_2}$, with the fixed charges located at the oxide/Si interface.
Fig. \ref{Fig:fig2_ppt_to_pdf_cropped}(e) shows the remote Coulomb scattering rate. The decrement of the scattering rate as the temperatures lowered resulted from the increment of the screening effect at low temperatures. The physics of ionized impurity scattering is similar to the remote Coulomb scattering. Ionized impurities form the scattering potential due to the Coulomb force. Compared with the remote Coulomb scattering, the charges causing the potential are located in the channel. 

The impurity concentration is treated as an assumed device parameter and is set to $\mathrm{5.0 \times 10^{11}}$\,$\mathrm{cm^{-2}}$, corresponding to an assumed uniform doping concentration of $\mathrm{1.0 \times 10^{18}}$\,$\mathrm{cm^{-3}}$ over the 5 nm Si channel thickness.
Fig. \ref{Fig:fig2_ppt_to_pdf_cropped}(f) shows the ionized impurity scattering rate. The decrement of the scattering rate as the temperatures lowered is also due to the increment of the screening effect at low temperatures.

\begin{figure}[!t]
\includegraphics[width=\linewidth, keepaspectratio]{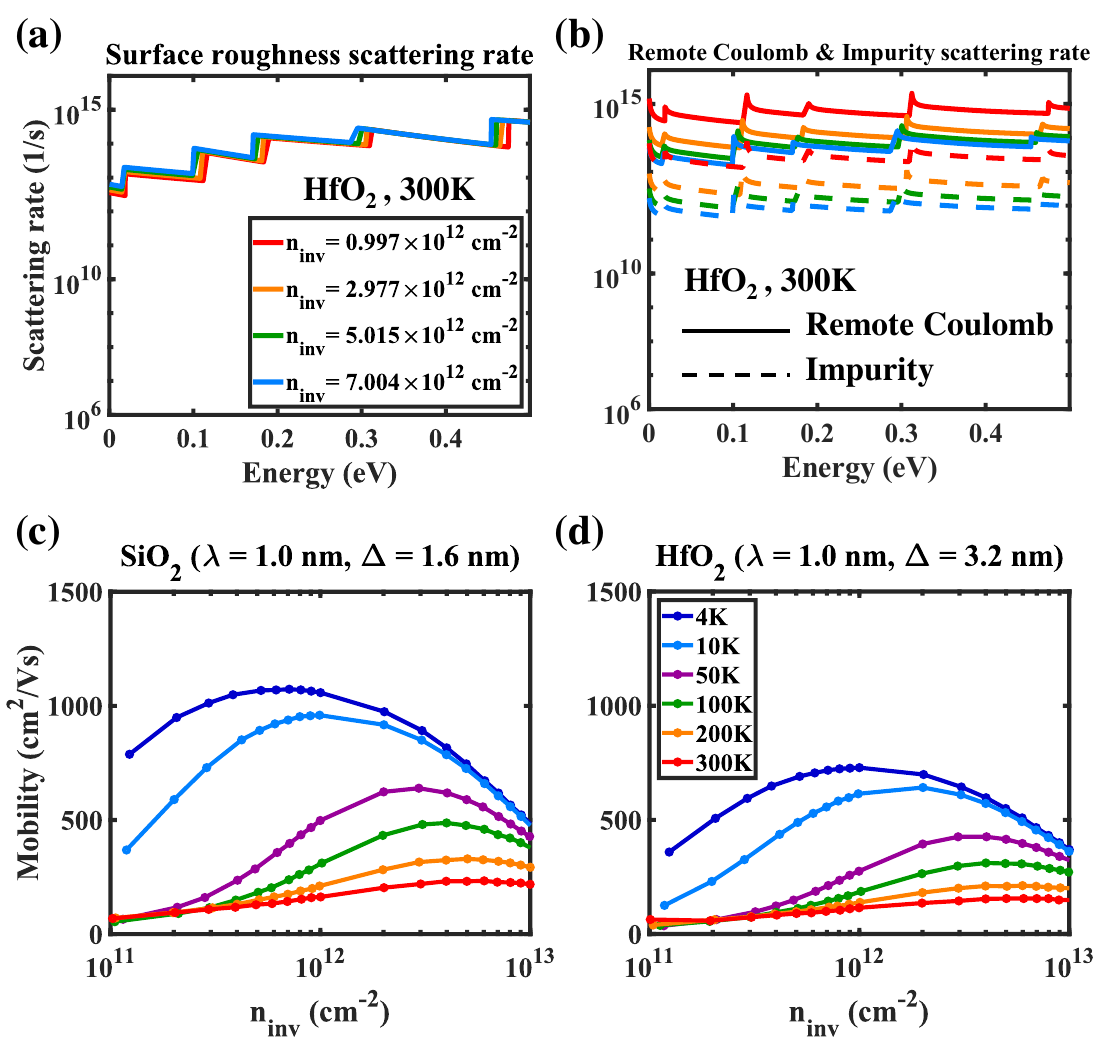}
\caption{\label{Fig:fig3_ppt_to_pdf_cropped}  
(a) Surface roughness scattering rate with different inversion layer concentrations ($\mathrm{n_{inv}}$). 
(b) Remote Coulomb scattering and ionized impurity scattering rates with different $\mathrm{n_{inv}}$. 
(c) and (d) Mobility with various $\mathrm{n_{inv}}$ when $\mathrm{SiO_2}$ and $\mathrm{HfO_2}$ are used as the dielectric, respectively. 
The same SRS, RCS, and IIS parameters are used in the scattering rate and mobility calculations.}
\end{figure}

If the device is under a different bias, the $\mathrm{n_{inv}}$ changes in the channel. Thus, the surface roughness scattering rate, remote Coulomb scattering rate, and ionized impurity scattering rate are influenced. 
The following calculations use the same empirical SRS and RCS parameters and the same assumed IIS concentration described above.
Figs. \ref{Fig:fig3_ppt_to_pdf_cropped}(a) and \ref{Fig:fig3_ppt_to_pdf_cropped}(b) are the results of surface roughness scattering rate, remote Coulomb scattering rate, and ionized impurity scattering rate with various $\mathrm{n_{inv}}$. Regarding the surface roughness scattering rate, the scattering rate enhanced as the inversion concentration increased. That is, a higher $\mathrm{n_{inv}}$ usually means more electrons are induced by gate bias and attracted to the Oxide/Si interfaces, which are more easily affected by the surface roughness. The rate decreases as the inversion layer concentration rises for remote Coulomb scattering and ionized impurity scattering rates due to carrier screening. Higher $\mathrm{n_{inv}}$ represents the stronger screening effect. Therefore, fewer scattering events happen at a higher $\mathrm{n_{inv}}$. Figs. \ref{Fig:fig3_ppt_to_pdf_cropped}(c) and \ref{Fig:fig3_ppt_to_pdf_cropped}(d) show the electron mobility with various $\mathrm{n_{inv}}$ under temperatures as the oxide is $\mathrm{SiO_2}$ or $\mathrm{HfO_2}$ (the same effective oxide thickness (EOT) is used). From the results, we can observe a peak mobility for $\mathrm{n_{inv}}$ near $\mathrm{10^{12}}$\,$\mathrm{cm^{-2}}$ to $\mathrm{5.0 \times 10^{12}}$\,$\mathrm{cm^{-2}}$ at different temperatures. This phenomenon results from the competition between surface roughness scattering and remote Coulomb scattering. The remote Coulomb scattering rate dominates at small $\mathrm{n_{inv}}$ due to weaker screening. The surface roughness scattering rate increases with increasing $\mathrm{n_{inv}}$ due to the higher electron distribution around the surface. 
The corresponding individual scattering rate components are provided in the Supplementary Information (Fig.~S1).

\begin{figure}[htbp]
\includegraphics[width=\linewidth, keepaspectratio]{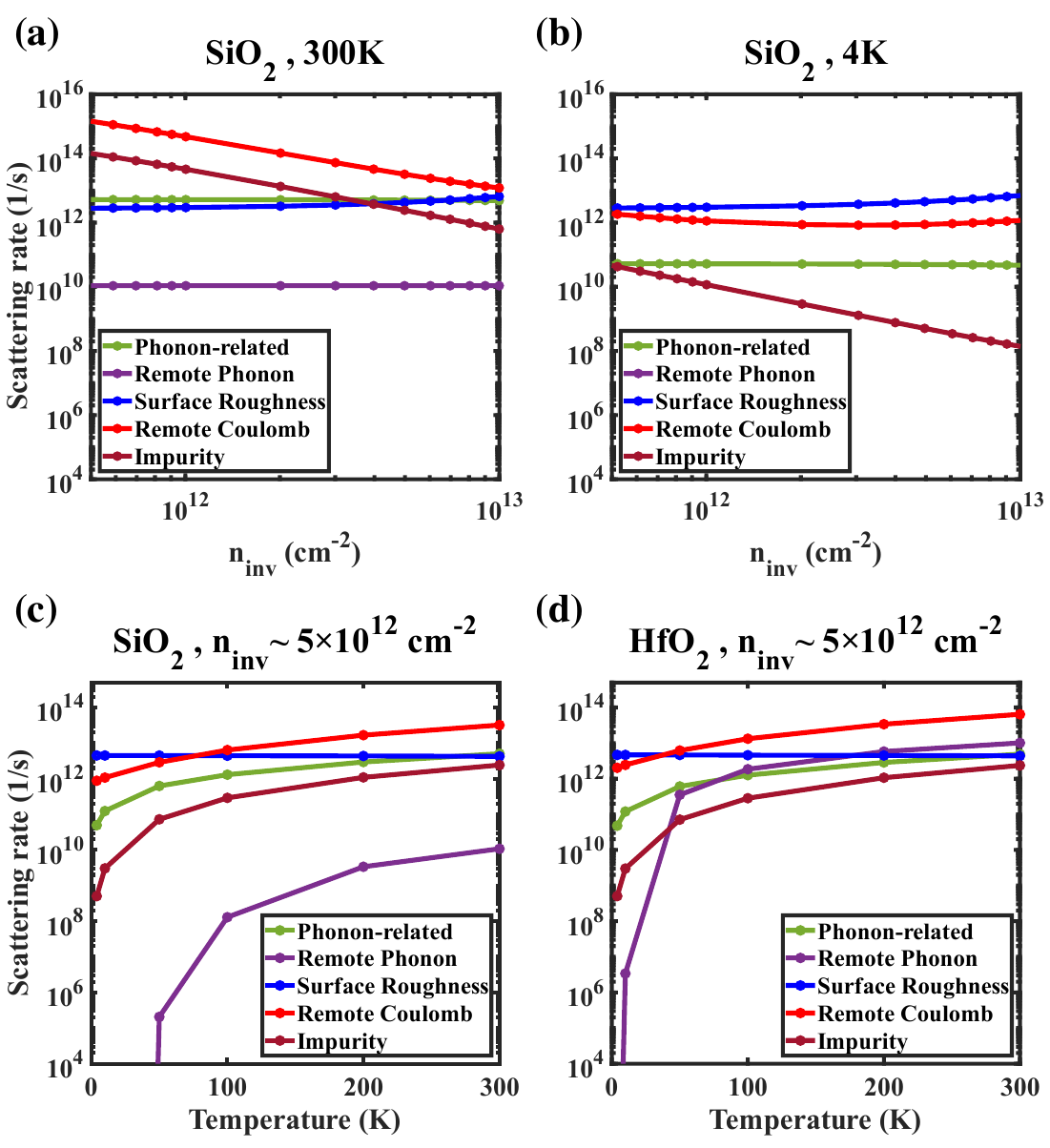}
\caption{\label{Fig:fig4_ppt_to_pdf_cropped}  
(a)$\sim$(b) Scattering rates (electron energy = 0.001 eV) with $\mathrm{n_{inv}}$ by using $\mathrm{SiO_2}$ as dielectric at 300K and 4K, respectively. (c)$\sim$(d) Scattering rates (electron energy = 0.001 eV) with temperatures and $\mathrm{SiO_2}$ and $\mathrm{HfO_2}$ are used as dielectric, respectively.  The same parameters in Fig. 2 are set.}
\end{figure}

Figs. \ref{Fig:fig4_ppt_to_pdf_cropped} (a) and \ref{Fig:fig4_ppt_to_pdf_cropped}(b) show the overall scattering rate changes with $\mathrm{n_{inv}}$ under 300K and 4K by using $\mathrm{SiO_2}$ as oxide. To begin with, at 4K, the phonon-related scattering rate magnitudes (green curve) are lower than those at 300K. Moreover, a competition between remote Coulomb scattering and surface roughness scattering (red and blue curves, respectively) can be observed with different $\mathrm{n_{inv}}$. 
In the 300K case, remote Coulomb scattering dominates at low $\mathrm{n_{inv}}$. With the rise of $\mathrm{n_{inv}}$, the surface roughness scattering rate increases, while remote Coulomb scattering becomes lower due to enhanced carrier screening. Therefore, at higher $\mathrm{n_{inv}}$
, the surface roughness scattering rate approaches the magnitude of the remote Coulomb scattering rate. 
Although the RCS--SRS crossover can be observed at 300K, the higher phonon-related scattering rate reduces its relative impact on the total scattering rate. Therefore, the corresponding mobility variation is less pronounced than that at 4K. After discussing the $\mathrm{n_{inv}}$-dependent scattering rates in the $\mathrm{SiO_2}$ case, Figs. \ref{Fig:fig4_ppt_to_pdf_cropped}(c) and \ref{Fig:fig4_ppt_to_pdf_cropped}(d) compare the temperature-dependent scattering rates at $\mathrm{n_{inv}}\sim\mathrm{5.0 \times 10^{12}}$\,$\mathrm{cm^{-2}}$ for the $\mathrm{SiO_2}$ and $\mathrm{HfO_2}$ cases, respectively. By comparing these two dielectric cases, a higher remote phonon scattering rate (purple curve) is observed in the HfO$_2$ case than in the SiO$_2$ case. For both dielectric cases, as the temperature decreases, phonon-related scattering decreases because of the reduced thermal phonon occupation. Remote Coulomb scattering and ionized impurity scattering also decrease due to the enhanced carrier screening at low temperatures, whereas surface roughness scattering remains nearly temperature independent.

\begin{figure}[!t]
\includegraphics[width=\linewidth, keepaspectratio]{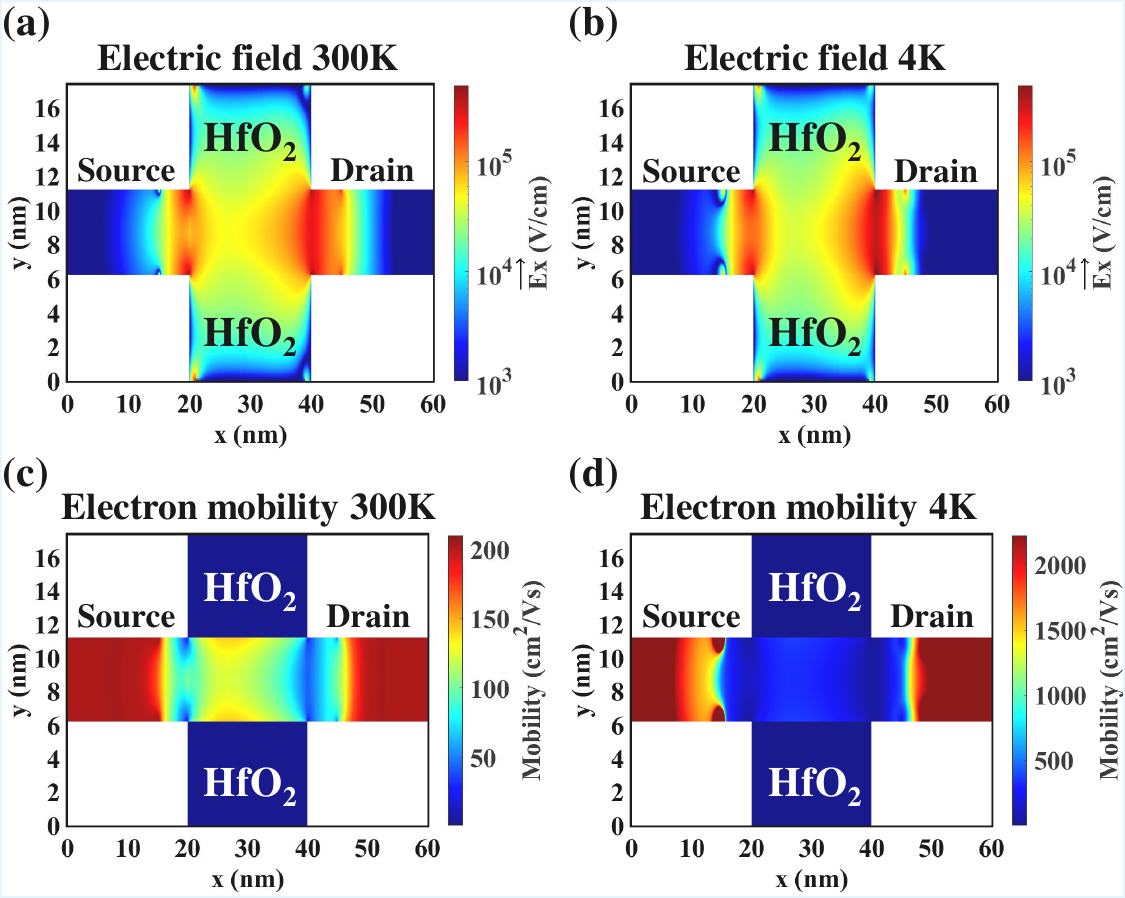}
\caption{\label{Fig:fig5_ppt_to_pdf_cropped}  (a) and (b) show the Electric field in the transport direction, $\vec{E}_{x}$, in the channel at 300K and 4K, respectively. (c) and (d) show the resulted effective high field mobility along the channel at 300K and 4K, respectively.}
\end{figure}

\begin{figure}[!t]
\includegraphics[width=\linewidth, keepaspectratio]{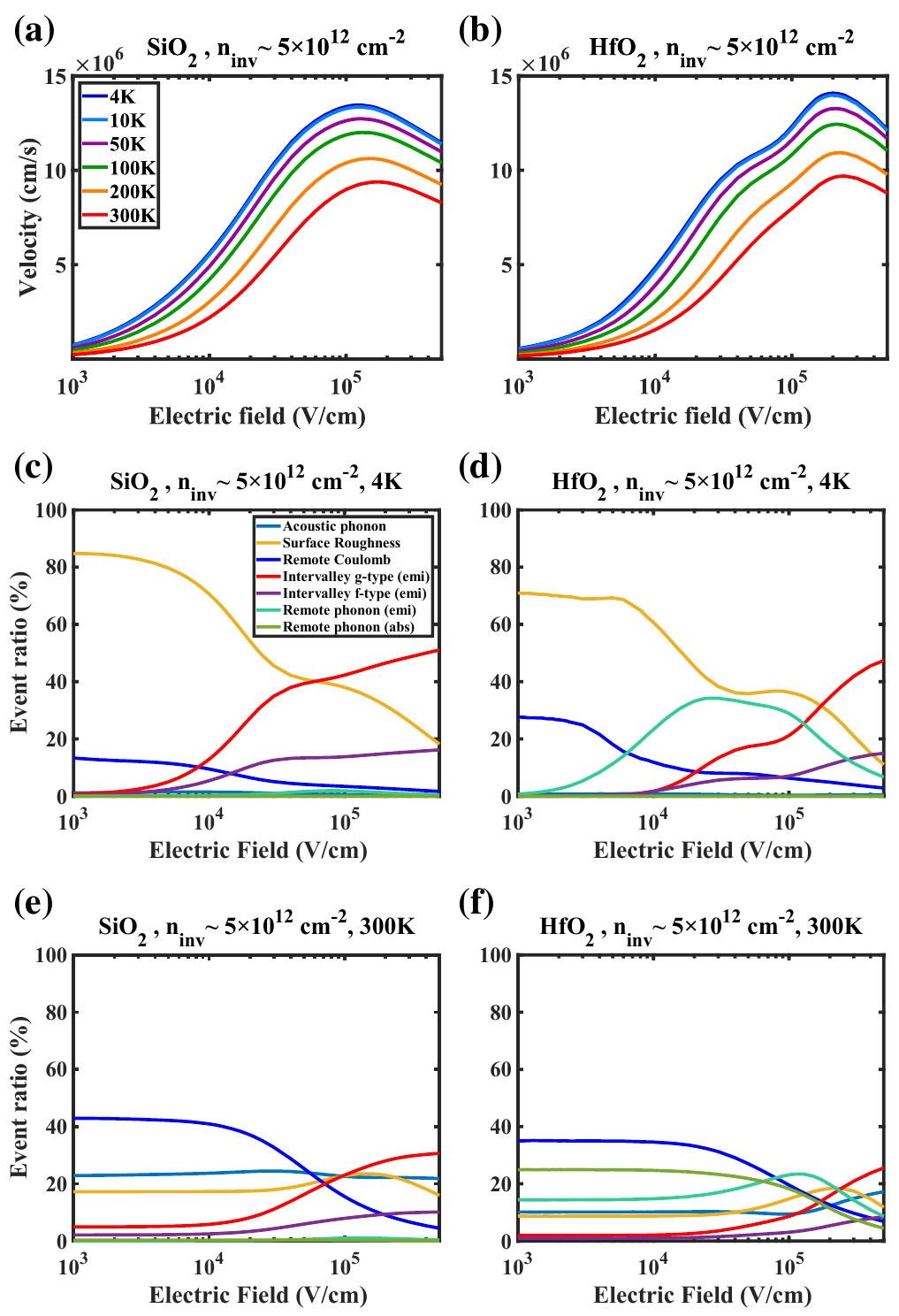}
\caption{\label{Fig:fig6_ppt_to_pdf_cropped}  
(a) and (b) represent the field-dependent velocity results for the $\mathrm{SiO_2}$ and $\mathrm{HfO_2}$ cases, respectively.
(c)--(f) Scattering event ratios of the main mechanisms, with 4K shown in (c) and (d) and 300K shown in (e) and (f) for the $\mathrm{SiO_2}$ and $\mathrm{HfO_2}$ cases, respectively.
The same parameters in Fig. 2 are set.
}
\end{figure}

The aforementioned discussion is based on the electron performance under a low electric field, $\vec{E}\parallel$, along the transport direction.
For transistor applications, especially in short gate-length cases, the electric field along the channel becomes very large, causing the electron velocity to reach saturation. Figures \ref{Fig:fig5_ppt_to_pdf_cropped}(a) and \ref{Fig:fig5_ppt_to_pdf_cropped}(b) show the electric field, $\vec{E}{x}$, along the channel. It can be observed that the electric field may exceed $10^{5}$ V/cm in the short-channel case. The effective mobility along the channel is therefore not governed by the low-field mobility but is instead influenced by the high-field velocity at each location. As a result, the current density is mainly determined by these high-field velocities, making it important to further investigate this effect.  Figs. \ref{Fig:fig6_ppt_to_pdf_cropped}(a) and \ref{Fig:fig6_ppt_to_pdf_cropped}(b) show the field-dependent velocity result by using $\mathrm{SiO_2}$ and $\mathrm{HfO_2}$ as oxide with $\mathrm{n_{inv}}\sim\mathrm{5.0 \times 10^{12}}$\,$\mathrm{cm^{-2}}$, and the electron performance in the higher electric field can be investigated. These plots show that the field-dependent velocity reaches a peak around or larger than $\mathrm{1.0 \times 10^{5}}$\,$\mathrm{V/cm}$,
depending on the temperature and dielectric material. In addition, a shoulder-like feature appears in the $\mathrm{HfO_2}$ case at low temperatures before the main high-field velocity peak, around $\mathrm{5.0 \times 10^{4}}$\,$\mathrm{V/cm}$. 
We record the scattering events to study which scattering mechanism causes this phenomenon. 
Figs. \ref{Fig:fig6_ppt_to_pdf_cropped}(c)--\ref{Fig:fig6_ppt_to_pdf_cropped}(f) are the event ratios of the main scattering mechanisms for the $\mathrm{SiO_2}$ and $\mathrm{HfO_2}$ cases at 4K and 300K. These event-ratio results show that
the peak around or above $\mathrm{\sim\mathrm{1.0 \times 10^{5}}}$\,$\mathrm{V/cm}$ resulted from the abrupt increment of g-type intervalley scattering (optical-phonon-related transition).
In this field range, carriers gain sufficient kinetic energy to emit g-type intervalley phonons with an energy of approximately 62 meV, resulting in enhanced g-type intervalley scattering. At lower electric fields preceding the main peak in the low-temperature $\mathrm{HfO_2}$ case, the remote phonon emission (emi) event ratio increases noticeably and is correlated with the shoulder-like feature in the velocity curve.
The carrier population fractions and average carrier kinetic energies at representative electric fields are examined in the Supplementary Information (Fig.~S2). At $\mathrm{5.0 \times 10^{4}}$\,$\mathrm{V/cm}$, the average carrier kinetic energies reach the energy range of several equivalent intervalley phonon modes listed in Table~S1, indicating that intervalley phonon emission can start to contribute in this field range. At $\mathrm{2.0 \times 10^{5}}$\,$\mathrm{V/cm}$, the higher carrier energies and broader X-subband population redistribution support the enhanced intervalley emission in the main high-field peak region.
The extracted low-field and high-field transport parameters at 4K are summarized in Table~\ref{tab:summary_4K}.

\begin{table}[htbp]
\caption{Summary of characteristic low-field and high-field transport parameters extracted at 4K for Si (110) confinement. Low-field mobility is evaluated at $E=1.0\times10^{3}$~V/cm as a function of $n_\mathrm{inv}$, while high-field parameters are extracted at $n_\mathrm{inv}\approx5.0\times10^{12}$~cm$^{-2}$.}
\label{tab:summary_4K}
\small
\centering
\renewcommand{\arraystretch}{1.2}
\begin{tabular*}{\columnwidth}{@{\extracolsep{\fill}}lcc}
\hline
\hline
\multicolumn{3}{c}{Low-field transport} \\
\hline
Parameter & SiO$_2$ & HfO$_2$ \\
\hline
$n_\mathrm{inv}^{\mathrm{peak}}$ (cm$^{-2}$) & $7\times10^{11}$ & $1\times10^{12}$ \\
$\mu_\mathrm{peak}$ (cm$^{2}$/V·s) & 1072.9 & 728.6 \\
\hline
\multicolumn{3}{c}{High-field transport} \\
\hline
Parameter & SiO$_2$ & HfO$_2$ \\
\hline
$E_\mathrm{crit}$ (V/cm) & $1.2\times10^{5}$ & $2.0\times10^{5}$ \\
$v_\mathrm{sat}$ (cm/s) & $1.35\times10^{7}$ & $1.41\times10^{7}$ \\
\hline
\hline
\end{tabular*}
\end{table}

Before summarizing the transport mechanisms, we further clarify the empirical parameter calibration used in this work. The phonon-related parameters, dielectric parameters, and band-structure parameters are kept fixed according to the literature values summarized in Table~S1. In contrast, the fixed-charge concentration $\mathrm{N}_{\mathrm{fix}}$ in the remote Coulomb scattering model and the surface roughness height $\Delta$ are treated as empirical fitting parameters. To calibrate these empirical parameters, single-parameter sweeps of $\mathrm{N}_{\mathrm{fix}}$ and $\Delta$ were examined at 300K while keeping the other model parameters fixed.

\begin{figure}[!t]
\includegraphics[width=\linewidth, keepaspectratio]{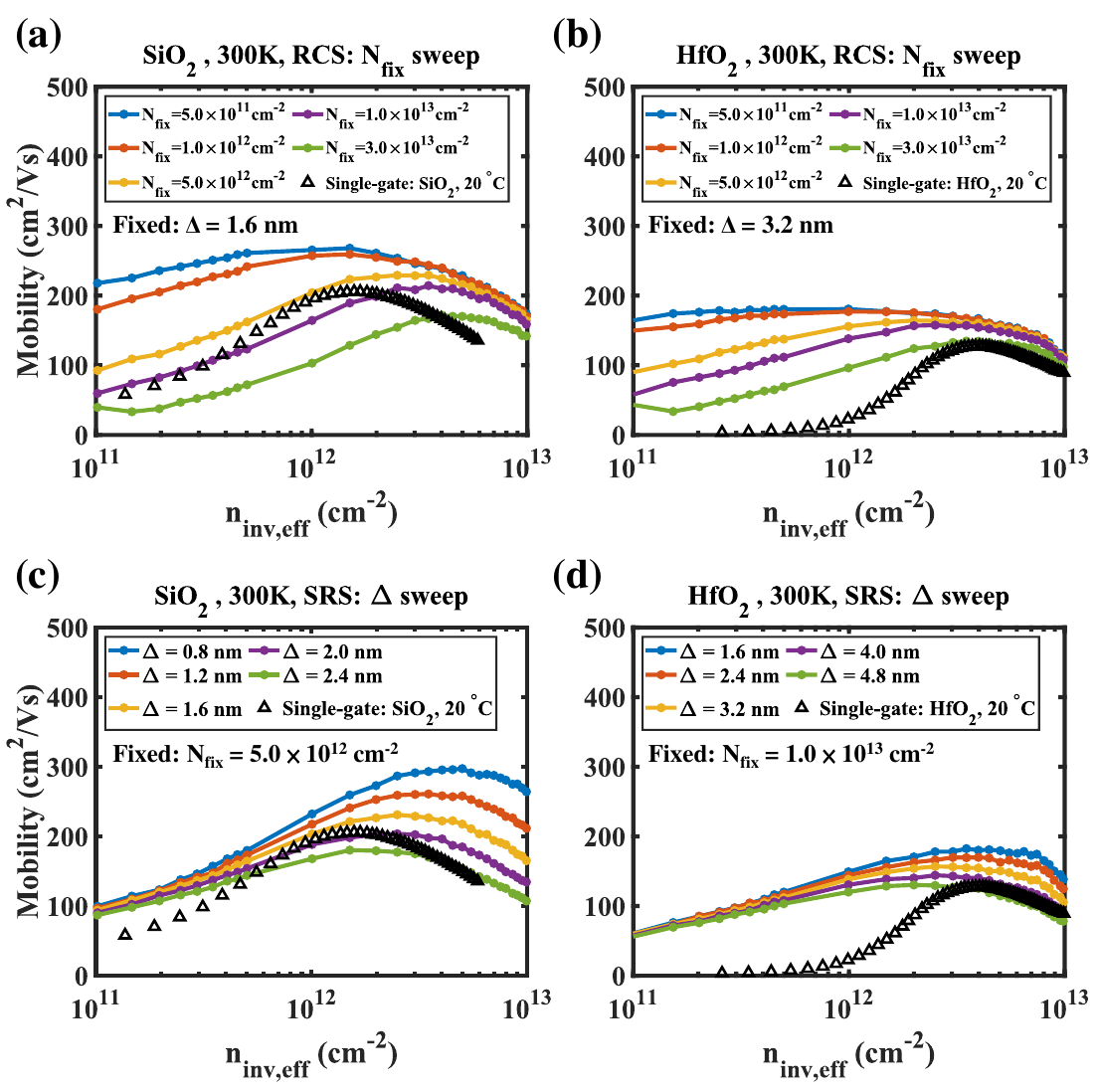}
\caption{\label{Fig:fig7_ppt_to_pdf_cropped}
Room-temperature literature benchmark and empirical-parameter calibration for the low-field mobility at 300K.
The $\mathrm{SiO_2}$ case is shown in (a) and (c), and the $\mathrm{HfO_2}$ case is shown in (b) and (d).
Panels (a) and (b) show single-parameter sweeps of the RCS fixed-charge concentration $\mathrm{N}_{\mathrm{fix}}$, while panels (c) and (d) show single-parameter sweeps of the SRS roughness height $\Delta$.
In each sweep, only the indicated empirical parameter is varied, and all other model parameters are fixed at the selected baseline values listed in Table~S1.
The black triangular markers are the experimental data from Mereu et al.~\cite{RN4}, and the colored curves are the simulated mobility curves plotted using $n_{\mathrm{inv,eff}}=n_{\mathrm{inv,total}}/2$.}
\end{figure}

Figure~\ref{Fig:fig7_ppt_to_pdf_cropped} shows the room-temperature mobility benchmark used for the empirical parameter calibration. The simulated mobility curves are compared with the Si(110)/$\langle 110\rangle$ n-MOSFET mobility data reported by Mereu et al.~\cite{RN4}. Since the present simulation employs a double-gate confinement geometry, whereas Ref.~\cite{RN4} corresponds to a conventional single-gate MOSFET structure, the simulated total inversion-layer concentration is converted to an effective value per interface, $n_{\mathrm{inv,eff}}=n_{\mathrm{inv,total}}/2$, for this benchmark comparison. This conversion is used to compare the characteristic density scale, including the mobility peak position, with the single-gate experimental data.

As shown in Fig.~\ref{Fig:fig7_ppt_to_pdf_cropped}, the simulation captures the representative room-temperature mobility scale and the experimentally observed trend that the mobility in the $\mathrm{HfO_2}$ case is lower than that in the $\mathrm{SiO_2}$ case. In the $\mathrm{N}_{\mathrm{fix}}$ sweeps shown in Figs.~\ref{Fig:fig7_ppt_to_pdf_cropped}(a) and \ref{Fig:fig7_ppt_to_pdf_cropped}(b), increasing $\mathrm{N}_{\mathrm{fix}}$ mainly suppresses the mobility in the low-to-intermediate $n_{\mathrm{inv,eff}}$ region, where carrier screening is relatively weaker and remote Coulomb scattering is more important. In the $\Delta$ sweeps shown in Figs.~\ref{Fig:fig7_ppt_to_pdf_cropped}(c) and \ref{Fig:fig7_ppt_to_pdf_cropped}(d), increasing $\Delta$ reduces the mobility more clearly at higher $n_{\mathrm{inv,eff}}$, where surface roughness scattering becomes more significant. Because the present work considers a different confinement geometry and gate-stack configuration from Ref.~\cite{RN4}, this comparison is intended as a literature-based benchmark for the overall room-temperature mobility level and trend, rather than as a one-to-one quantitative fit to the experimental device.

Based on this comparison, one baseline parameter set was selected for each dielectric case and then used consistently in the subsequent scattering rate, mobility, and high-field velocity calculations. The selected values are $\mathrm{N}_{\mathrm{fix}}=5.0\times10^{12}$~cm$^{-2}$ and $\Delta=1.6$~nm for the $\mathrm{SiO_2}$ case, and $\mathrm{N}_{\mathrm{fix}}=1.0\times10^{13}$~cm$^{-2}$ and $\Delta=3.2$~nm for the $\mathrm{HfO_2}$ case, with $\lambda=1.0$~nm for both cases. The cryogenic mobility enhancement is guided by the low-temperature FinFET measurements reported by Han et al.~\cite{RN32}, where the nMOS effective electron mobility at 2.95K was shown to be more than four times its room-temperature value. The representative 4K sensitivity to $\mathrm{N}_{\mathrm{fix}}$ and $\Delta$ is further shown for the $\mathrm{SiO_2}$ case in the Supplementary Information (Fig.~S3).

\section{Conclusion}
This study investigated electron transport properties in silicon (110) confinement systems under cryogenic conditions using a multi-valley Monte Carlo simulation approach. The simulation framework accounted for various scattering mechanisms, including acoustic phonon, optical phonon, surface roughness, remote Coulomb, and ionized impurity scattering. Our results revealed the temperature dependence of these scattering rates, emphasizing the dominant role of surface roughness and Coulomb scattering at low temperatures where phonon-related scattering diminishes.

We also explored the impact of different dielectric materials ($\mathrm{SiO_2}$ and $\mathrm{HfO_2}$) on electron mobility. Our findings demonstrated that while high-$\kappa$ materials like $\mathrm{HfO_2}$ can enhance gate control and device performance, they also increase remote phonon scattering rates compared to $\mathrm{SiO_2}$. This trade-off suggests the need for optimized dielectric selection when designing cryogenic silicon devices.

Further analysis indicated that electron mobility behavior is influenced significantly by the inversion layer concentration ($\mathrm{n_{inv}}$), with a clear competition between surface roughness and remote Coulomb scattering mechanisms. The results highlight that at lower inversion concentrations, remote Coulomb scattering dominates, but as $\mathrm{n_{inv}}$ increases, surface roughness scattering becomes more significant, reducing mobility. These insights provide a deeper understanding of the balance between scattering mechanisms in cryogenic environments.

Finally, the velocity versus electric field in the transport direction is studied. It revealed that optical phonon emission plays a crucial role in intervalley scattering, influencing the electron velocity profiles, especially under higher electric fields. These results validate the importance of considering both material and operational conditions when designing advanced silicon-based devices for low-temperature applications.
\section{Supplementary Information}
\textbf{Supplementary Information} is available at
\href{https://doi.org/10.1088/1361-6463/ae7b4f/data1}{https://doi.org/10.1088/1361-6463/ae7b4f/data1}.

\begin{acknowledgments}
This work is supported by the National Science and Technology Council under grant No. 112-2221-E-002-214-MY3, 113-2124-M-002-013-MY3, 113-2218-E-002-034 and 114-2622-8-002-016.   Prof. Yuh-Renn Wu is also supported by the Leap fellowship of the Foundation for the Advancement of Outstanding Scholarship.
\end{acknowledgments}

\nocite{*}
\bibliography{aipsamp}

\end{document}